\documentclass[12pt,aps,prd,preprint,tightenlines,superscriptaddress,
   showpacs,nofootinbib]{revtex4}
\usepackage{hyperref}
\usepackage{graphicx}
\usepackage{epsfig}
\usepackage{amsmath, amsfonts,euscript,bbm}
\usepackage{ifthen}
\usepackage{graphicx}

\newcommand{\roughly}[1]{\mathrel{\raise.3ex\hbox{$#1$\kern-0.85em
\lower1ex\hbox{$\sim$}}}}

\def\be{\begin{equation}}
\def\beq\begin{equation}
\def\ee{\end{equation}}
\def\bea{\begin{eqnarray}}
\def\eea{\end{eqnarray}}

\def\eqref#1{(\ref{#1})}

\def\UV{{\scriptscriptstyle U \kern-.1emV}}
\def\IR{{\scriptscriptstyle I\kern-.18em R}}

\def\gsim{\ \rlap{\raise 3pt \hbox{$>$}}{\lower 3pt \hbox{$\sim$}}\ }
\def\lsim{\ \rlap{\raise 3pt \hbox{$<$}}{\lower 3pt \hbox{$\sim$}}\ }

\begin{document}

\thispagestyle{empty}

\preprint{UCI-HEP-TR-2012-20}

\title{ Effective Theories of Gamma-ray Lines from Dark Matter Annihilation \vspace*{0.5cm}
}

\author{Arvind Rajaraman}
\affiliation{Department of Physics and Astronomy,
University of California, Irvine, CA 92697, USA \vspace*{0.5cm}}

\author{Tim M.P. Tait}
\affiliation{Department of Physics and Astronomy,
University of California, Irvine, CA 92697, USA \vspace*{0.5cm}}

\author{Alexander M. Wijangco}
\affiliation{Department of Physics and Astronomy,
University of California, Irvine, CA 92697, USA \vspace*{0.5cm}}

\date{\today}


\begin{abstract}
We explore theories of dark matter in which dark matter annihilations produce mono-energetic gamma rays (``lines") in
the context of effective field theory, which captures the physics for cases in which the particles mediating the interaction
are somewhat heavier than the dark matter particle itself.  Building on earlier work, we explore the generic signature
resulting from $SU(2) \times U(1)$ gauge invariance that two (or more) lines are generically expected, and determine the
expected relative intensities, including the possibility of interference between operators.
\end{abstract}

\maketitle

\section{Introduction}

While the existence of  dark matter is now secure,  its nature remains elusive.
Many experiments are searching for
evidence of  non-gravitational dark matter interactions  through  direct detection of its scattering
off heavy nuclei, or by direct production in colliders.
Yet another approach to search for dark matter indirectly is by looking for signals
produced by its annihilation to Standard model (SM) particles.
In particular, the Universe is transparent to $\sim 100$~GeV energy
gamma rays on galactic distance scales, which allows one to use their distribution in the sky as well as
in energy as handles to try to sift the signal from the (often poorly understood) astrophysical backgrounds.

Among the most striking potential signals one can imagine from dark matter annihilation is
a mono-energetic ``line" of gamma rays.  Such a process occurs when two (non relativistic) dark matter particles
annihilate into a two-body final state, one of which is a photon.  The most canonical of such signals
would be annihilation into two photons, whose energies are expected to be very close to the mass
of the dark matter particle.  While such a process is usually suppressed
compared to the continuum of gamma rays that result from
dark matter annihilations into charged or hadronic particles,
the signature is distinctive and difficult for more conventional astrophysics to mimic.

In fact, recent analysis of data obtained by the  Fermi-LAT collaboration \cite{Abdo:2010nc,Fermi:2012}
has found tentative indications of such
a line at an energy of about 130 GeV \cite{Bringmann:2012vr,Weniger:2012tx,Su:2012ft}, originating from 
close to the
galactic center \cite{Bringmann:2012vr,Weniger:2012tx,Su:2012ft}.  Such a signal is tantalizing, and the
presence of what may be a fainter secondary line in the data whose energy is consistent with annihilation into
a $\gamma Z$ final state \cite{Su:2012ft,Rajaraman:2012db} lends some credence to an interpretation in terms
of dark matter annihilation.  On the other hand, searches for signs of a signal in targets away from the 
galactic center
have yielded results which are confusing at best 
\cite{Tempel:2012ey,Hektor:2012kc,Su:2012zg,Hooper:2012qc,Hektor:2012jc},
there are significant limits on a continuum signal associated with the regions of the sky where the line appears
brightest \cite{Cohen:2012me},
and most seriously, there seems to be a hint of a feature in photons arriving from the direction of the 
Earth's limb
\cite{Su:2012ft} raising the possibility that the feature in the data is the result of a subtle instrumental
effect \cite{Whiteson:2012hr,Hektor:2012ev,Finkbeiner:2012ez}.  Perhaps less likely, the signal could also 
correspond
to more prosaic astrophysical processes
\cite{Profumo:2012tr,Boyarsky:2012ca,Aharonian:2012cs}. Despite these potential issues, the
feature at 130 GeV  is very interesting and worthy of investigation.

In this article, we examine dark matter annihilations as a source of multiple lines,
using the powerful language of
effective field theory developed in Ref.~\cite{Rajaraman:2012db}.  While we are inspired by the feature 
currently observed
in the Fermi data, we will be more concerned with the generic systematics of multiple line signals, which can
be applied both to the currently observed feature and to future searches.

\section{Effective Field Theory}
\label{sec:ops}

We will take the dark matter (denoted by $\chi$) to be a singlet under all SM interactions, which implies 
that it couples to the
electroweak gauge bosons (including the photon) through higher dimensional operators that result from
integrating out electroweakly charged massive fields.
We work at the level of this effective field theory containing the
dark matter and the SM itself.  We impose a $Z_2$ symmetry under which the dark matter is odd and the SM is 
even in
order to insure that the dark matter is stable.  We are interested in operators containing at least one photon, 
so as to
result in an observable gamma ray line signal.  The operators are organized by the energy dimension of the 
field content,
since one generically expects operators corresponding to higher dimensions to be less relevant in low energy 
processes,
being more suppressed by the masses of the heavy particles that were integrated out to produce them.

This last issue raises an important question -- does one expect that effective field theory can capture the physics of dark
matter annihilation into photons at all?  Like any effective theory, our theory is valid at very low momentum 
transfer, but
fails to capture the physics of high energy processes, for which the complete theory in the ultra-violet is 
required.  For
(non-relativistic) dark matter annihilation, the characteristic momentum transfer is of order the mass of the dark matter itself,
and so this assumption boils down to the requirement that the particles mediating the interaction between the dark matter
and electroweak bosons are heavier than $m_\chi$.  Since such mediator particles must be charged, their masses
are bounded in general to be $\gsim 100$~GeV by LEP (or more by the LHC if they are stable on collider time scales
\cite{Chatrchyan:2012sp}).
However, in many theories the loop process
connecting the dark matter to the weak bosons contains a mixture of SM as well as heavy mediator particles.  For example,
the line signal resulting from dark matter whose primary interaction is with SM light quarks is considered in
Ref.~\cite{Goodman:2010qn}.

\begin{center}
\begin{table}[t]
\begin{tabular}{|c|c|c|}
\hline
\multicolumn{3}{|c|}{Dimension 6 Operators} \\
\hline
B1+B2 & $\frac{1}{\Lambda^2_{B1}}\chi\chi^* B_{\mu\nu}B^{\mu\nu}
+\frac{1}{\Lambda^2_{B2}}\chi\chi^* W^a_{\mu\nu}W^{a\mu\nu}$ & $\gamma \gamma$, $\gamma Z$ \\
B3+B4 & $\frac{1}{\Lambda^2_{B3}}\chi\chi^* B_{\mu\nu}\tilde{B}^{\mu\nu}
+\frac{1}{\Lambda^2_{B4}}\chi\chi^* W^a_{\mu\nu}\tilde{W}^{a\mu\nu}$ & $\gamma \gamma$, $\gamma Z$ \\
\hline
\multicolumn{3}{|c|}{Dimension 8 Operators} \\
\hline
D1+D2 & $\frac{1}{\Lambda^4_{D1}}(\chi\partial_\mu\chi^*-\chi^*\partial_\mu\chi)B^{\mu\alpha}\Phi^\dag D_\alpha\Phi
+\frac{1}{\Lambda^4_{D2}}(\chi\partial_\mu\chi^*-\chi^*\partial_\mu\chi)\Phi^\dag W^{\mu\alpha}_aT^aD_\alpha\Phi$
&  \\
D3+D4 & $\frac{1}{\Lambda^4_{D3}}(\chi\partial_\mu\chi^*-\chi^*\partial_\mu\chi)
\tilde{B}^{\mu\alpha}\Phi^\dag D_\alpha\Phi
+\frac{1}{\Lambda^4_{D4}}(\chi\partial_\mu\chi^*-\chi^*\partial_\mu\chi)\Phi^\dag \tilde{W}^{\mu\alpha}_aT^aD_\alpha\Phi$
& \\
\hline
\end{tabular}
\caption{List of effective interactions for complex scalar dark matter and the type of line signals ($\gamma \gamma$,
$\gamma Z$, and/or $\gamma h$) that they produce.}
\label{Operatorlist-s}
\end{table}
\end{center}

Nonetheless, even if there are light SM particles participating, the structure remains tightly constrained by $SU(2) \times U(1)$
electroweak gauge invariance.  The presence of multiple lines thus remains generic and (provided the non-SM heavy mediators
are sufficiently heavy compared to $m_\chi$), the relative rates of various line processes such as $\gamma \gamma$
and $\gamma Z$ are not likely to show large deviations from an effective theory description, although the precise mapping of the
EFT coefficients to the UV theory parameters becomes more murky.

Our effective field theory is constructed as the Standard Model, plus a dark matter particle $\chi$ which we allow to be either
a complex scalar or Dirac fermion.  The real and Majorana cases are simply related to our results.  The interactions of
interest contain at least one $SU(2)$ $W_3^\mu$ or hyper charge $B^\mu$ gauge field, which will become a photon after
rotating to the mass eigenbasis,
\begin{align}
B_\mu=&A_\mu\cos\theta_W-Z_\mu\sin\theta_W\\\nonumber
W^3_\mu=&A_\mu\sin\theta_W+Z_\mu\cos\theta_W
\end{align}
where $A_\mu$ and $Z_\mu$ are the photon and $Z$ boson fields respectively, and
$\theta_W$ is the electroweak mixing angle.

\begin{center}
\begin{table}[t]
\begin{tabular}{|c|c|c|}
\hline
\multicolumn{3}{|c|}{Dimension 5 Operators} \\
\hline
A1+A2 & $\frac{1}{\Lambda_{A1}}\bar{\chi}\gamma^{\mu\nu}\chi B_{\mu\nu}
+\frac{1}{\Lambda_{A2}}\bar{\chi}\gamma^{\mu\nu}\chi \tilde{B}_{\mu\nu}$ &  $\gamma \gamma$, $\gamma Z$ \\
\hline
\multicolumn{3}{|c|}{Dimension 7 Operators} \\
\hline
C1+C2 & $\frac{1}{\Lambda^3_{C1}}\bar{\chi}\chi B_{\mu\nu}B^{\mu\nu}
+\frac{1}{\Lambda^3_{C2}}\bar{\chi}\chi W^a_{\mu\nu}W^{a\mu\nu}$ & \\
C3 + C4& $\frac{1}{\Lambda^3_{C3}}\bar{\chi}\chi B_{\mu\nu}\tilde{B}^{\mu\nu}
+\frac{1}{\Lambda^3_{C4}}\bar{\chi}\chi W^a_{\mu\nu}\tilde{W}^{a\mu\nu}$ & \\
C5+C6 & $\frac{1}{\Lambda^3_{C5}}\bar{\chi}\gamma^5\chi B_{\mu\nu}B^{\mu\nu}
+\frac{1}{\Lambda^3_{C6}}\bar{\chi}\gamma^5\chi W^a_{\mu\nu}W^{a\mu\nu}$ & $\gamma \gamma$, $\gamma Z$ \\
C7 +C8& $\frac{1}{\Lambda^3_{C7}}\bar{\chi}\gamma^5\chi B_{\mu\nu}\tilde{B}^{\mu\nu}
+\frac{1}{\Lambda^3_{C8}}\bar{\chi}\gamma^5\chi W^a_{\mu\nu}\tilde{W}^{a\mu\nu}$ & $\gamma \gamma$, $\gamma Z$ \\
C9+C10 & $\frac{1}{\Lambda^3_{C9}}\bar{\chi}\gamma^{\mu\nu}\chi B_{\mu\alpha}\tilde{B}^{\alpha\nu}
+\frac{1}{\Lambda^3_{C10}}\bar{\chi}\gamma^{\mu\nu}\chi W^a_{\mu\alpha}\tilde{W}^{a\alpha\nu}$ &  $\gamma Z$\\
C11+C12 & $ \frac{1}{\Lambda^3_{C11}}\bar{\chi}\gamma^{\mu\nu}\chi B_{\mu\nu}|\Phi|^2
+\frac{1}{\Lambda^3_{C12}}\bar{\chi}\gamma^{\mu\nu}\chi\Phi^\dag W^a_{\mu\nu}T^a\Phi$ & $\gamma h$ \\
C13+C14 & $\frac{1}{\Lambda^3_{C13}}\bar{\chi}\gamma^{\mu\nu}\chi \tilde{B}_{\mu\nu}|\Phi|^2
+\frac{1}{\Lambda^3_{C14}}\bar{\chi}\gamma^{\mu\nu}\chi\Phi^\dag \tilde{W}^a_{\mu\nu}T^a\Phi$ & $\gamma h$ \\
\hline
\multicolumn{3}{|c|}{Dimension 8 Operators} \\
\hline
D5+D6  & $\frac{1}{\Lambda^4_{D5}}\bar{\chi}\gamma_\mu\chi B^{\mu\alpha}\Phi^\dag D_\alpha\Phi
+\frac{1}{\Lambda^4_{D6}}\bar{\chi}\gamma_\mu\chi \Phi^\dag W^{\mu\alpha}_aT^aD_\alpha\Phi$ &
$\gamma Z$, $\gamma h$ \\
D7+D8 & $\frac{1}{\Lambda^4_{D7}}\bar{\chi}\gamma_\mu\chi \tilde{B}^{\mu\alpha}\Phi^\dag D_\alpha\Phi
+\frac{1}{\Lambda^4_{D8}}\bar{\chi}\gamma_\mu\chi \Phi^\dag \tilde{W}^{\mu\alpha}_aT^aD_\alpha\Phi$ &
$\gamma Z$, $\gamma h$ \\
D9+D10 & $\frac{1}{\Lambda^4_{D9}}\bar{\chi}\gamma_\mu\gamma_5\chi B^{\mu\alpha}\Phi^\dag D_\alpha\Phi
+\frac{1}{\Lambda^4_{D10}}\bar{\chi}\gamma_\mu\gamma_5\chi \Phi^\dag W^{\mu\alpha}_aT^aD_\alpha\Phi$ &
$\gamma Z$ \\
D11+D12 & $\frac{1}{\Lambda^4_{D11}}\bar{\chi}\gamma_\mu\gamma_5\chi \tilde{B}^{\mu\alpha}\Phi^\dag D_\alpha\Phi
+\frac{1}{\Lambda^4_{D12}}\bar{\chi}\gamma_\mu\gamma_5\chi \Phi^\dag \tilde{W}^{\mu\alpha}_aT^aD_\alpha\Phi$ &
$\gamma Z$ \\
\hline
\end{tabular}
\caption{List of effective interactions for Dirac fermion dark matter and the type of line signals ($\gamma \gamma$,
$\gamma Z$, and/or $\gamma h$) that they produce.}
\label{Operatorlist-f}
\end{table}
\end{center}

We consider the effective vertices shown in Tables~\ref{Operatorlist-s} and \ref{Operatorlist-f} (for scalar and fermionic
dark matter, respectively), which are built out of the dark matter fields, the field strengths $W_a^{\mu \nu}$ and
$B^{\mu \nu}$ (and their duals $\widetilde{W}$ and $\widetilde{B}$),
the SM Higgs doublet $\Phi$, and its covariant derivative
\begin{equation}
D_\alpha \Phi = \partial_\alpha \Phi
-i g_2 T_a W^a_\alpha \Phi - i\frac{1}{2} g_1 B_\alpha \Phi~,
\end{equation}
where $T^a$ are the generators of the doublet representation of $SU(2)$,
and $g_1$ and $g_2$ are gauge couplings. We define $\gamma^{\mu\nu}\equiv[\gamma^\mu,\gamma^\nu]$.
For our purposes, it will be sufficient to work in the unitary gauge, for which
we may take,
\bea
\Phi \rightarrow \frac{1}{\sqrt{2}} \left(
\begin{array}{c}
0 \\
V + h(x)
\end{array}
\right)~,
\eea
where $V \simeq 246$~GeV is the Higgs vacuum expectation value and
$h(x)$ is the physical Higgs field.

This set of interactions is complete up to terms of dimension 8.  A similar
list was studied, including bounds from indirect and direct detection and
LHC searches, in Ref.~\cite{Cotta:2012nj}.

\section{Line Cross Sections}

Since dark matter is expected to be highly non-relativistic (with
velocity dispersion $v \sim 10^{-3}$) in the galactic halo, dark matter
annihilation into photons may be simplified as an expansion in $v^2$.
We retain only the leading ($v$-independent) terms.  In this limit,
the operators C1 -- C4 and D1 -- D4 lead to vanishing cross sections, and
thus are unlikely to lead to any observable line signal.
Operators A1 and A2 (which
correspond to magnetic/electric dipole moments for the dark matter)
are strongly constrained by direct detection \cite{Fortin:2011hv},
and thus also unlikely to contribute to a large line
signal\footnote{An inelastically scattering dark matter particle with
dipole interactions can evade direct detection constraints and might even
explain the DAMA signal, see
Refs.~\cite{Goodman:2010qn,Chang:2010en,Kumar:2011iy,Weiner:2012cb,Weiner:2012gm}.}.
We leave consideration of all of these unpromising cases for future work.

We will denote $p_1, p_2$ to be the incoming dark matter particle momenta,
$p_3$ will be a photon, and $p_4$ is either another
photon, $Z$ boson, or higgs boson.
 The differential cross section is written
\begin{align}
\frac{d\sigma}{d\Omega}
=&\frac{E_3}{256\pi^2~E^3~v} \overline{|{\cal M} |^2}
\end{align}
where $E = m_\chi + {\cal O}(v^2)$ is the energy of each dark matter particle,
$v$ is the dark matter velocity, and $E_3 = |\vec{p}_3|$ is the energy of
the outgoing line photon.  $|{\cal M} |^2$ is the matrix element
${\cal M}$ averaged over
initial dark matter spins (if any) and summed over final state particle spins.

\subsection{Dimension 6 Operators}

Operators B1 -- B4 and C1 -- C18 all have the form
$X F^{\mu\nu} F_{\mu\nu}$ or $X F^{\mu\nu} \widetilde{F}_{\mu\nu}$, where $F^{\mu \nu} = B^{\mu \nu}$ or
$W^{\mu \nu}$.
For the $X F^{\mu\nu} F_{\mu\nu}$ operators,
the matrix element will have the form
${\cal M}=-2Y(p_3\cdot p_4 \epsilon_3\cdot \epsilon_4)$
where $Y$ is whatever the Feynman rules of $X$ yield, and depends on the spin of the dark matter.
If $p_4$ corresponds to another photon, we find
\begin{equation}
\displaystyle\sum_{\epsilon_3,\epsilon_4}|{\cal M}|^2
=16YY^\dag(p_3\cdot p_4)^2~.
\end{equation}
To obtain $\overline{| {\cal M}|^2}$, one averages this result over the dark matter
spin states.
In the case where $p_4$ corresponds to a massive gauge boson, we find:
\begin{align}
\displaystyle\sum_{\epsilon_3,\epsilon_4}|{\cal M}|^2
=&12YY^\dag(p_3\cdot p_4)^2~.
\end{align}

For an operator of the form
$X F^{\mu\nu}\tilde{F}_{\mu\nu}$
the matrix element will have the form:
${\cal M}= Y(p^\mu_3 \epsilon^\nu_3-p^\nu_3 \epsilon^\mu_3)(p^\rho_4 \epsilon^\sigma_4-p^\sigma_4 \epsilon^\rho_4)\epsilon_{\mu\nu\rho\sigma}$
Squaring this yields:
\begin{equation}
\displaystyle\sum_{\epsilon_3,\epsilon_4}|{\cal M}|^2=32YY^\dag(p_3\cdot p_4)^2
~.
\end{equation}

For the (complex) scalar dark matter dimension six operators, $Y Y^\dag=\frac{1}{\Lambda^2}$ and the average
over spin is trivial.  The operators B1 and B2 interfere with one another, but are separate from B3 and B4.
The resulting cross sections are:

\bea
|v|\sigma_{B1,2}(\chi\chi^* \rightarrow\gamma\gamma) & = & \frac{2m_\chi^2}{\pi}\left(\frac{\cos^4\theta_W}
{\Lambda^4_{B1}}+\frac{2\cos^2\theta_W\sin^2\theta_W}{\Lambda^2_{B1}\Lambda^2_{B2}}
+\frac{\sin^4\theta_W}{\Lambda^4_{B2}}\right) ~, \\
|v|\sigma_{B1,2}(\chi\chi^* \rightarrow\gamma Z) & = & \frac{3\cos^2\theta_W\sin^2\theta_W(4m_\chi^2-m^2_Z)^3}{64\pi m_\chi^4}
\left(\frac{1}{\Lambda^4_{B1}}-\frac{2}{\Lambda^2_{B1}\Lambda^2_{B2}}
+\frac{1}{\Lambda^4_{B2}}\right) ~,
\eea
and
\bea
|v|\sigma_{B3,4}(\chi\chi^* \rightarrow\gamma\gamma) &=& \frac{4m_\chi^2}{\pi}\left(\frac{\cos^4\theta_W}
{\Lambda^4_{B3}}+\frac{2\cos^2\theta_W\sin^2\theta_W}{\Lambda^2_{B3}\Lambda^2_{B4}}
+\frac{\sin^4\theta_W}{\Lambda^4_{B4}}\right) ~,\\
|v|\sigma_{B3,4}(\chi\chi^* \rightarrow\gamma Z) &=& \frac{\cos^2\theta_W\sin^2\theta_W(4m_\chi^2-m^2_z)^3}
{8\pi m_\chi^4}\left(\frac{1}{\Lambda^4_{B3}}-\frac{2}{\Lambda^2_{B3}\Lambda^2_{B4}}+\frac{1}{\Lambda^4_{B4}}
\right) ~,
\eea
respectively.

\subsection{Dimension 7 Operators}

A Dirac fermion can annihilate into $\gamma \gamma$ and $\gamma Z$ through the dimension seven operators
C5 -- C8 (recall that C1 -- C4 vanish at zero velocity).  The matrix elements are identical to B1 -- B4 as far as the
final state, and the only difference is the average over initial WIMP spins.  The resulting cross sections are,
\bea
|v|\sigma_{C5,6}(\chi \bar{\chi}\rightarrow\gamma\gamma) &=& \frac{4m^4_\chi}{\pi}
\left(\frac{\cos^4\theta_W}{\Lambda^6_{C5}}+\frac{2\cos^2\theta_W\sin^2\theta_W}
{\Lambda^3_{C5}\Lambda^3_{C6}}+\frac{\sin^4\theta_W}{\Lambda^6_{C6}}\right)~, \\
|v|\sigma_{C5,6}(\chi\bar{\chi}\rightarrow\gamma Z) &=& \frac{3(4m_\chi^2-m^2_Z)^3\cos^2\theta_W\sin^2\theta_W}{32\pi m_\chi^2}
\left(\frac{1}{\Lambda^6_{C5}}-\frac{2}{\Lambda^3_{C5}
\Lambda^3_{C6}}+\frac{1}{\Lambda^6_{C6}}\right)~,
\eea
and
\bea
|v|\sigma_{C7,8}(\chi\bar{\chi}\rightarrow\gamma\gamma) &=& \frac{8m^4_\chi}{\pi}
\left(\frac{\cos^4\theta_W}{\Lambda^6_{C7}}+\frac{2\cos^2\theta_W\sin^2\theta_W}
{\Lambda^3_{C7}\Lambda^3_{C8}}+\frac{\sin^4\theta_W}{\Lambda^6_{C8}}\right)~, \\
|v|\sigma_{C7,8} (\chi\bar{\chi}\rightarrow\gamma Z) &=& \frac{(4m_\chi^2-m^2_Z)^3
(\cos^2\theta_W\sin^2\theta_W)}{4\pi m_\chi^2}\left(\frac{1}{\Lambda^6_{C7}}
-\frac{2}{\Lambda^3_{C7}\Lambda^3_{C8}}+\frac{1}{\Lambda^6_{C8}}\right)~.
\eea

The remaining dimension seven operators lead to single lines.  For C9 and C10, the
antisymmetry of $\gamma^{\mu\nu}$ forces the $\chi\chi\rightarrow\gamma\gamma$ cross section to vanish identically,
leaving only a $\gamma Z$ line:
\begin{align}
|v|\sigma_{C9,10}(\chi\bar{\chi}\rightarrow\gamma Z) =&
\frac{(4m^2_\chi-m^2_Z)^3(4m^2_\chi+m^2_Z)\cos^2\theta_W\sin^2\theta_W}{16m^4_\chi\pi}
\left(\frac{1}{\Lambda^6_{C9}}-\frac{2}{\Lambda^3_{C9}\Lambda^3_{C10}}
+\frac{1}{\Lambda^6_{C10}}\right)~.
\end{align}
Whereas
operators C11 -- C14 result in a single $\gamma h$ line,
\bea
|v|\sigma_{C11,12}(\chi\bar{\chi}\rightarrow\gamma h)
&=& \frac{(4m^2_\chi-m^2_h)^3V^2}{64m^4_\chi\pi}\left(\frac{\cos^2\theta_W}
{\Lambda^6_{C11}}-\frac{\cos\theta_W\sin\theta_W}{\Lambda^3_{C11}\Lambda^3_{C12}}
+\frac{\sin^2\theta_W}{4\Lambda^6_{C12}}\right)~,
\eea
and
\bea
|v|\sigma_{C13,14}(\chi\bar{\chi}\rightarrow\gamma h)
&=& \frac{(4m^2_\chi-m^2_h)^3V^2}{16m^4_\chi\pi}\left(\frac{\cos^2\theta_W}
{\Lambda^6_{C13}}-\frac{\cos\theta_W\sin\theta_W}{\Lambda^3_{C13}\Lambda^3_{C14}}
+\frac{\sin^2\theta_W}{4\Lambda^6_{C14}}\right)~.
\eea

\subsection{Dimension 8 Operators}

Dimension eight operators could in principle contribute to line signals from scalar dark matter, but in practice these
operators lead to cross sections which vanish in the zero velocity limit.   Thus, we limit our discussion to the case
where the dark matter is a Dirac fermion, for which there are potentially both $\gamma Z$ and $\gamma h$ final states.
For D5 -- D8, we have two sets of interfering operators,
\begin{align}
|v|\sigma_{D5,6}(\chi \bar{\chi} \rightarrow\gamma Z)
=&\frac{(4m_\chi^2-m_Z^2)^3(4m_\chi^2+m_Z^2)V^4(g_2\cos\theta_W+g_1\sin\theta_W)^2}
{4096\pi m_\chi^4m_Z^2}\\\nonumber
&\times\left(\frac{\cos^2\theta_W}
{\Lambda_{D5}^8}-\frac{\cos\theta_W\sin\theta_W}{\Lambda_{D5}^4\Lambda_{D6}^4}
+\frac{\sin^2\theta_W}{4\Lambda_{D6}^8}\right)~, \\
|v|\sigma_{D5,6}(\chi \bar{\chi} \rightarrow\gamma h)=&\frac{(4m_\chi^2-m_h^2)^3V^2}{1024m_\chi^2\pi}\left(\frac{\cos^2\theta_W}
{\Lambda_{D5}^8}-\frac{\cos\theta_W\sin\theta_W}{\Lambda_{D5}^4\Lambda_{D6}^4}
+\frac{\sin^2\theta_W}{4\Lambda_{D6}^8}\right)~,
\end{align}
and
\begin{align}
|v|\sigma_{D7,8}(\chi \bar{\chi}\rightarrow\gamma Z)=&
\frac{(4m_\chi^2-m_Z^2)^3(24m_\chi^2+m_Z^2)V^4(g_2\cos\theta_W+g_1\sin\theta_W)^2}
{1024\pi m_\chi^4m_Z^2}\\\nonumber
&\times\left(\frac{\cos^2\theta_W}{\Lambda_{D7}^8}-\frac{\cos\theta_W\sin\theta_W}
{\Lambda_{D7}^4\Lambda_{D8}^4}+\frac{\sin^2\theta_W}{4\Lambda_{D8}^8}\right)~, \\
|v|\sigma_{D7,8}(\chi \bar{\chi}\rightarrow\gamma h)=&
\frac{(4m_\chi^2-m_h^2)^3V^2}{256\pi m_\chi^2}\left(\frac{\cos^2\theta_W}
{\Lambda_{D7}^8}-\frac{\cos\theta_W\sin\theta_W}
{\Lambda_{D7}^4\Lambda_{D8}^4}+\frac{\sin^2\theta_W}
{4\Lambda_{D8}^8}\right)~.
\end{align}

For operators D9 -- D12, the $\gamma h$ line vanishes in the limit of zero velocity, leaving a single bright
$\gamma Z$ line from each set of operators.  The cross sections are,
\begin{align}
|v|\sigma_{D9,10}(\chi \bar{\chi}\rightarrow\gamma Z)=&\frac{(4m^2_\chi-m^2_Z)^3V^4(g_2\cos\theta_W+g_1\sin\theta_W)^2}{16384m_\chi^4\pi}\\\nonumber
&\times\left(\frac{\cos^2\theta_W}{\Lambda_{D9}^8}-\frac{\cos\theta_W\sin\theta_W}{\Lambda_{D9}^4\Lambda_{D10}^4}+\frac{\sin^2\theta_W}{4\Lambda_{D10}^8}\right)~,
\end{align}
and
\begin{align}
|v|\sigma_{D11,12}(\chi \bar{\chi}\rightarrow\gamma Z)=&\frac{(4m^2_\chi-m^2_Z)^3V^4(g_2\cos\theta_W+g_1\sin\theta_W)^2}{4096m_\chi^4\pi}\\\nonumber
&\times\left(\frac{\cos^2\theta_W}{\Lambda_{D11}^8}-\frac{\cos\theta_W\sin\theta_W}{\Lambda_{D11}^4\Lambda_{D12}^4}+\frac{\sin^2\theta_W}{4\Lambda_{D12}^8}\right)~.
\end{align}

\section{Summary}

We are now in a position to summarize the various possible annihilation modes for each operator class. The processes
resulting from each operator which are not suppressed by the dark matter velocity are listed in in the third column
of Tables~\ref{Operatorlist-s} and \ref{Operatorlist-f}.
As is evident from the table, any operator which can produce a $\gamma \gamma$ line will
(modulo interference between two operators) also result in a $\gamma Z$
one, whereas some of the higher dimension operators are able to produce $\gamma Z$ or $\gamma h$ lines in isolation.
Of course, a specific UV theory of dark matter may result in more than one operator being turned on.  Typically one expects
that relevant operators of the lowest dimension will dominate the size of each line with corrections from higher
order terms being controlled by $m_\chi / \Lambda_i$ to the appropriate power.

Our results are suggestive of new ways to interpret the results of line searches.  Given a choice of dark matter mass
and now that the Large Hadron Collider has measured the Higgs boson mass, the energy of each line is determined,
\bea
E_{\gamma \gamma} & = & m_\chi \\
E_{\gamma Z} & = & \frac{4m_\chi^2-m_Z^2}{4m_\chi} \\
E_{\gamma h} & = & \frac{4m_\chi^2-m_h^2}{4m_\chi}
\eea
where it should be clear that the energy in each case refers to the energy of the final state photon, and the label
applies to the process which produced the gamma ray.  Since multiple
lines are a fairly generic feature, it would be interesting to recast single line searches into searches for multiple lines
based on a given value of $m_\chi$.  For example, a search for lines related to a scalar dark matter particle could search
simultaneously for two lines with energies $E_{\gamma \gamma}$ and $E_{\gamma Z}$ based on operators B1 and B2.
At each putative dark matter mass, a bound can be placed in the $\Lambda_{B1}$-$\Lambda_{B2}$ plane.

Alternately,
if one has a particular UV theory in mind such that either one operator or the other (or some linear combination
with a fixed ratio) is generated, one can improve the sensitivity by searching for two lines at correlated energies with a fixed
intensity ratio for the two.  In Table~\ref{annihilationModes5}, we list, for each operator, the strength of the
first (lowest energy) and second line implied by each set of operators.
For convenience, we have introduced the short hand notation:
\bea
f_1(\Lambda_1,\Lambda_2,n) &\equiv& \left(\frac{\cos^4\theta_W}{\Lambda_1^{2n}}
+\frac{2\cos^2\theta_W\sin^2\theta_W}{\Lambda_1^n\Lambda_2^n }+\frac{\sin^4\theta_W}{\Lambda_2^{2n}}\right)~, \\
f_2(\Lambda_1,\Lambda_2,n) &\equiv& \cos^2\theta_W\sin^2\theta_W\left(\frac{1}{\Lambda_1^{2n}}
-\frac{2}{\Lambda_1^n\Lambda_2^n }+\frac{1}{\Lambda_2^{2n}}\right)~, \\
f_3(\Lambda_1,\Lambda_2,n,m) &\equiv& \left(\frac{\cos^2\theta_W}{\Lambda_1^{2n}}-
\frac{\cos\theta_W\sin\theta_W}{\Lambda_1^n\Lambda_2^n }+\frac{\sin^2\theta_W}{4\Lambda_2^{2n}}\right)
(g_2\cos\theta_W+g_1\sin\theta_W)^m~.
\eea
The operator groups D5+D6 and D7+D8 each predict a fixed ratio between the two lines, regardless of the
specifics of the relative coefficients of the operators within each category.  The ratios are:
\bea
\frac{|v|\sigma_{D5,6}(\chi\chi\rightarrow\gamma Z)}{|v|\sigma_{D5,6}(\chi\chi\rightarrow\gamma h)}
&=&
\left(\frac{4m_\chi^2-m_Z^2}{4m_\chi^2-m_h^2}\right)^3\frac{(4m_\chi^2+m_Z^2)V^2(g_2\cos\theta_W
+g_1\sin\theta_W)^2}
{4 ~m_\chi^2 ~ m_Z^2}~, \\
\frac{|v|\sigma_{D7,8}(\chi\chi\rightarrow\gamma Z)}{|v|\sigma_{D7,8}(\chi\chi\rightarrow\gamma h)} &=&
\left(\frac{4m_\chi^2-m_Z^2}{4m_\chi^2-m_h^2}\right)^3\frac{(24m_\chi^2+m_Z^2)V^2(g_2\cos\theta_W
+g_1\sin\theta_W)^2}{4~m_\chi^2 ~ m_Z^2}~.
\eea

\begin{center}
\begin{table}
\centering
	\begin{tabular}{|c|cc|}
		\hline
		Operator & First Line & Second Line \\
		\hline
		B1+B2 & $\frac{3(4m_\chi^2-m^2_Z)^3}{64\pi m_\chi^4}f_2(\Lambda_{B1},\Lambda_{B2},2)$ & $\frac{2m_\chi^2}{\pi}f_1(\Lambda_{B1},\Lambda_{B2},2)$ \\
		B3+B4 & $\frac{(4m_\chi^2-m^2_z)^3}{8\pi m_\chi^4}f_2(\Lambda_{B3},\Lambda_{B4},2)$ & $\frac{4m_\chi^2}{\pi}f_1(\Lambda_{B7},\Lambda_{B8},2)$ \\
		C5+C6 & $\frac{3(4m_\chi^2-m^2_Z)^3}{32\pi m_\chi^2}f_2(\Lambda_{C5},\Lambda_{C6},3)$ & $\frac{4m^4_\chi}{\pi}f_1(\Lambda_{C5},\Lambda_{C6},3)$ \\
		C7+C8 & $\frac{(4m_\chi^2-m^2_Z)^3}{4\pi m_\chi^2}f_2(\Lambda_{C7},\Lambda_{C8},3)$ & $\frac{8m^4_\chi}{\pi}f_1(\Lambda_{C7},\Lambda_{C8},3)$ \\
		C9+C10 & $\frac{(4m^2_\chi-m^2_Z)^3(4m^2_\chi+m^2_Z)}{16m^4_\chi\pi}f_2(\Lambda_{C9},\Lambda_{C10},3)$ & N/A \\
		C11+C12 & $\frac{(4m^2_\chi-m^2_h)^3V^2}{64m^4_\chi\pi}f_3(\Lambda_{C11},\Lambda_{C12},3,0)$ & N/A \\
		C13+C14 & $\frac{(4m^2_\chi-m^2_h)^3V^2}{16m^4_\chi\pi}f_3(\Lambda_{C13},\Lambda_{C14},3,0)$ & N/A \\
		D5+D6 & $\frac{(4m_\chi^2-m_h^2)^3V^2}{1024m_\chi^2\pi}f_3(\Lambda_{D5},\Lambda_{D6},4,0)$ & 
        $\frac{(4m_\chi^2-m_Z^2)^3(4m_\chi^2+m_Z^2)V^4}{4096\pi m_\chi^4m_Z^2}f_3(\Lambda_{D5},\Lambda_{D6},4,2)$ \\
		D7+D8 & $\frac{(4m_\chi^2-m_h^2)^3V^2}{256\pi m_\chi^2}f_3(\Lambda_{D7},\Lambda_{D8},4,0)$ & 
        $\frac{(4m_\chi^2-m_Z^2)^3(24m_\chi^2+m_Z^2)V^4}{1024\pi m_\chi^4m_Z^2}f_3(\Lambda_{D7},\Lambda_{D8},4,2)$ \\
		D9+D10 & $\frac{(4m^2_\chi-m^2_Z)^3V^4}{16384m_\chi^4\pi}f_3(\Lambda_{D9},\Lambda_{D10},4,2)$ & N/A \\
		D11+D12 & $\frac{(4m^2_\chi-m^2_Z)^3V^4}{4096m_\chi^4\pi}f_3(\Lambda_{D11},\Lambda_{D12},4,2)$ & N/A \\
		\hline
	\end{tabular}
\caption{The strength of the first and second (when applicable) gamma ray line signals for each operator described in the
text.}
\label{annihilationModes5}
\end{table}
\end{center}

\section{Outlook}

Gamma ray line searches make up a crucial part of searches for the indirect detection of dark matter.  We have studied
using the tools of effective field theory the generic multi-line signatures of dark matter annihilation.  While our specific
analytic results apply only in the case where the particles mediating the interactions between the dark matter and
photons are much heavier than the dark matter itself, the central result that there are multiple lines, and their relative
intensities, are the consequence of $SU(2) \times U(1)$ gauge invariance, and thus rather generic.

We have examined the set of lowest operators which can contribute to $\gamma \gamma$, $\gamma Z$, and $\gamma h$
lines, for dark matter which is a complex scalar or Dirac fermion.  Our results suggest that an interesting extension of the
current suite of searches for photon lines at gamma ray telescopes would include the simultaneous search for two lines
at fixed relative energies.  Such a search should improve the sensitivity to specific UV theories of dark matter in many cases,
which fix the ratio between the interfering operators of a given dimensionality.  Should a set of lines be discovered, the
energies and relative intensities of the set provide key information as to the possible responsible operators, and thus the
first clues as to the nature of the dark matter responsible.

\acknowledgments

TMPT acknowledges the Aspen Center for Physics (supported by the NSF grant PHY-1066293)
where part of this work was completed.
and the theory group of SLAC National Lab for their support of his many visits.
The work of AR and TMPT is supported in part by NSF grants PHY-0970173 and PHY-0970171.

\end{document}